# Nanoplasmon-enabled macroscopic thermal management

Gustav Edman Jonsson[1], Vladimir Miljkovic[1] and Alexandre Dmitriev[1]*

[1] Department of Applied Physics, Chalmers University of Technology, Göteborg 41296, Sweden.

* Corresponding author: Alexandre Dmitriev, +46-31-7725177, alexd@chalmers.se

**In numerous applications of energy harvesting via transformation of light into heat the focus recently shifted towards highly absorptive materials featuring nanoplasmons[1, 2, 3, 4, 5]. It is currently established that noble metals-based absorptive plasmonic platforms deliver significant light-capturing capability and can be viewed as super-absorbers of optical radiation[6, 7]. However, direct experimental evidence of plasmon-enabled** *macroscopic* **temperature increase that would result from these efficient absorptive properties is scarce. Here we derive a general quantitative method of characterizing light-capturing properties of a given heat-generating absorptive layer by macroscopic thermal imaging. We further monitor macroscopic areas that are homogeneously heated by several degrees with plasmon nanostructures that occupy a mere 8% of the surface, leaving it essentially transparent and evidencing significant heat generation capability of nanoplasmon-enabled light capture. This has a direct bearing to thermophotovoltaics and other applications where thermal management is crucial.**

Practical thermal platforms for solar energy harvesting and manipulation of thermal radiation are necessarily highly absorbing, but also affordable and scalable. Nanoplasmonic metamaterial absorbers offer a route towards low material consumption and compact designs while maintaining large optical cross-sections[7]. They exploit the collective oscillations of electrons in metallic nanostructures – localized plasmons – for the highly efficient coupling of light. In the same time, individual plasmon nanostructures in these materials act collectively as an effective absorbing layer, i.e., a metasurface. Whereas up to now the focus is on intricate design of nanoplasmonic metamaterials units, often by advanced electromagnetic simulations and highly demanding nanofabrication methods such as electron-beam lithography, recent advances in nanofabrication have led to simple and thereby cheap large-scale manufacturing methods of light absorbing plasmonic metasurfaces[8, 9]. These approaches utilize amorphous arrangements of individual nanoplasmonic structures (meta-atoms) with simple geometries, which as a whole develop desired collective absorptive properties. When developing realistic light absorbing nanoplasmonic systems, optimization primarily focuses on the size, shape and material composition of the individual particles. While local plasmon-enabled thermal gradients, confined to the nanoscale, are well-documented in these systems[10, 11, 12, 13, 14, 15, 16], large-scale (i. e., macroscopic) thermal effects due to plasmons are not properly addressed. And in the wake of the development of macroscopic plasmon metamaterials absorbers that aim, for example, at wafer-scale thermophotovoltaics, the comprehensive experimental evidence of such thermal effects is required.

The primary aim when developing absorbing plasmonic architectures is to enhance the non-radiative decay of the excited nanoplasmons. This is typically done by reducing the size of plasmonic nanoparticles or by changing the dielectric function of the material[17]. Although noble metal nanoparticles of Au or Ag are widely used in plasmon metamaterials due to their well-pronounced resonances in the visible and near-IR, they might not be the best choice as absorbing platforms. Under white light illumination plasmonic nanostructures having markedly lower but instead spectrally broader absorption may be equal to or even surpassing Au-based superabsorbers.

Here we utilize simple large-area bottom-up nanofabrication method – hole mask colloidal lithography[9] – to prepare absorbing plasmonic metasurfaces, composed of amorphous arrays of Ni or Au nanodisks that are 20 nm high and nominally 110 nm in diameter (Figure 1a). Figure 1b shows the absorption in these two types of nanoplasmonic metasurfaces, along with that of a carbon film of equal thickness that we use as the absorption benchmark. Au nanodisks layer shows strong and spectrally narrow absorption peak at the plasmon resonance. The Ni nanodisks system displays broader and generally lower absorption as a result of the shorter plasmon lifetime that is due to the strong electron-lattice coupling in Ni.

Although optical spectroscopy measurements give straightforward quantification of light absorption in these metasurfaces, they understandably do not address the crucial information on the ability to generate elevated temperature at a macroscale due to such absorption. With the current study we put forward a new modality of characterizing macroscopic absorbing metasurfaces – that is, via their temperature increase upon illumination. We naturally consider that all absorbed light eventually transforms into heat in these systems. With this measuring modality we illuminate the surface with the solar simulator to benefit from potential omnichromatic absorption of the practical metamaterial absorber and track the resulting surface temperature with a thermal camera (Figure 1c, further experimental details can be found in Methods section). Due to invariant thermal radiation emissivity properties of the substrates, we arrange the measuring system in a way that the ultra-thin absorbing plasmon metasurface is positioned at the backside of the sample and glass slide is facing light source and the thermal camera (c.f., Fig. 1c). The temperature is further obtained by averaging the camera reading from the entire sample (typically, 4 cm$^2$) first without the illumination (2 min), then with light on and reaching temperature saturation (about 3 min), and finally after light is off tracking the cooling-off kinetics (extra 4 min). The result for the amorphous array of Au nanodisks as an absorber is pictured in Figure 1d, with vertical axis presenting the difference to the temperature in the dark (room temperature). At this point one clearly marks the macroscopic temperature effect due to the presence of plasmon metasurface and overall sample temperature increase up to 5-6 degrees. We note here that temperature evolution in Fig. 1d concerns the entire system – that is, plasmon metasurface, glass slide and the surrounding air altogether. Specifically, at the point of temperature saturation, a thermal equilibrium exists between these three media, and the temperature gradients are established within the glass substrate – the hottest being plasmon metasurface layer.

To quantify the actual absorbed power we employ a simple one-dimensional numerical model: we consider a single domain consisting of the glass substrate with heat dissipating at the boundaries. Once a heat source is placed at one boundary, the temperature is extracted at the opposite side. By varying the power of the simulated heat source we find the best fit to the measured temperature response (see Methods for details). Figure 2a schematically outlines the general procedure of extracting the absorbed power with the given system (the actual fit to experimental data is shown) and cross-referencing it to the measured optical absorption. The absorbed powers for three systems extracted both from the temperature data (light grey) and optical absorption (dark grey) are presented in Figure 2b. Here we compare two nanoplasmonic platforms (amorphous array of Au or Ni nanodisks) and carbon film. For completeness we also plot the ratio of the two (green bars). The latter visually confirms the consistency between the thermally derived absorption and the optically measured one in all three systems, even though the absolute values of the thermally derived absorbance are underestimated as compared to their optical counterparts due to over-simplicity of the employed model.

We further establish that the quantitative aspects of the absorptive properties of a given thin functional layer, in principle, of unknown composition and morphology can be directly extracted from the thermal data. For example, comparing absorptive Au metamaterials surface with carbon film, one concludes that the carbon film absorbs 4.5 times more than Au metamaterial. In our case this is also directly verified by the measured optical absorbance in all these layers. Excitingly, if the morphology of the absorptive layer is known (i.e., film thickness and/or nanostructures size and coverage – i.e., the overall content of the absorptive material is quantified) – straightforward comparison of the heat generation ability of these ultrathin macroscopic functional layers can further be made. In our case, normalizing the thermal power on materials coverage both for nanoplasmonic metasurfaces and carbon film, we estimate nanoplasmonic systems to deliver 3.4 times higher absorption / thermal power per surface unit given that they have a surface coverage of 8 per cent.

Now that we established the method to correlate absorption and thermal effects, we turn back to the plasmon-enabled heat-generation platforms to mark the possible route towards an optimal plasmonic 'heater'. As mentioned earlier, gold- and silver-based plasmonic nanoarchitectures typically display strong and spectrally well-defined region of light absorption that is linked to the plasmon resonance. By contrast, metals like Ni display markedly broader plasmon feature in absorption due to higher ohmic losses. We use the latter property to design a simple plasmonic temperature system that favorably compares to structurally identical one, but made of Au. Namely, we use presented method to compare the absorbed power in amorphous arrays of Au and Ni nanoellipses. Figures 3a and 3b compare longitudinal and transversal optical absorption of Au and Ni nanoellipses, respectively (the overall array morphology is depicted in Fig. 3b). Even though the peak optical absorption in Ni system is obviously lower, significant spectral broadening of the absorption in Ni makes the total area under the peak – which in this sense corresponds to the actual absorption over the entire spectral range – noticeably larger. This larger area directly reports itself in the thermally measured data (Figure 3d), where Ni nanoellipses generate almost 40 % more heat. This signals that Ni-based plasmonic absorbers would potentially deliver appreciably larger and spectrally broader heat-generation capability for practical applications.

This brings forward the potential of other 'non-conventional' plasmonic materials – such as Pd, Pt, Co and others – as plasmon-enabled thermal metamaterial platforms for solar energy harvesting[18]. They typically feature very moderate absolute absorption in a broad spectral range (c.f., Fig. 1b) that also comfortably covers solar radiation spectrum. The former leaves macroscopic areas that are functionalized with thermal plasmon nanostructures essentially transparent. In the same time, these nanostructures provide superior heat-generation ability. With the simultaneous access to design and affordable fabrication of truly large-scale thermal plasmonic metasurfaces from these materials, the concepts of plasmon-enabled heating might spill into applications of thermal surfaces, for example, in energy-efficient technologies for construction and industrial design.

Another exciting prospect is in thermal plasmonics of nanoferromagnets for the manipulation of magnetic order. Indeed, plasmon-induced temperature effects are currently actively explored with magnetoplasmonics [19, 20] With the current work we show that Ni as efficient nanoplasmon heat antenna [21, 22] has substantial heat-generation ability and thus might find direct use in promoting strong nanoscale temperature gradients for effective magnetization manipulation.

Summarizing, we devised a new modality to comprehensively characterize the absorbing and heat-generating ability of various ultra-thin absorbers, most importantly including plasmon-enabled macroscopic absorbing metasurfaces.  It is also very appealing that the method allows for quantitative comparison of fundamentally different absorbing systems. Further, we used the method to mark the possible route towards potentially high-efficient plasmon absorbers that utilize Ni rather than Au. All these powerful features open up new possibilities for the reliable evaluation and practical implementation of ultra-thin metamaterial superabsorbers in thermophotovoltaics, energy-efficient design, thermal magnetization control and other applications where thermal management is important.

**Acknowledgments:**

We acknowledge Raja Sellappan (Chalmers) for providing the samples with carbon films. We also acknowledge Henrik Frederiksen (Chalmers) for his support in nanofabrication. This work is supported by the Swedish Foundation for Strategic Research (SSF) through RMA08-0109.

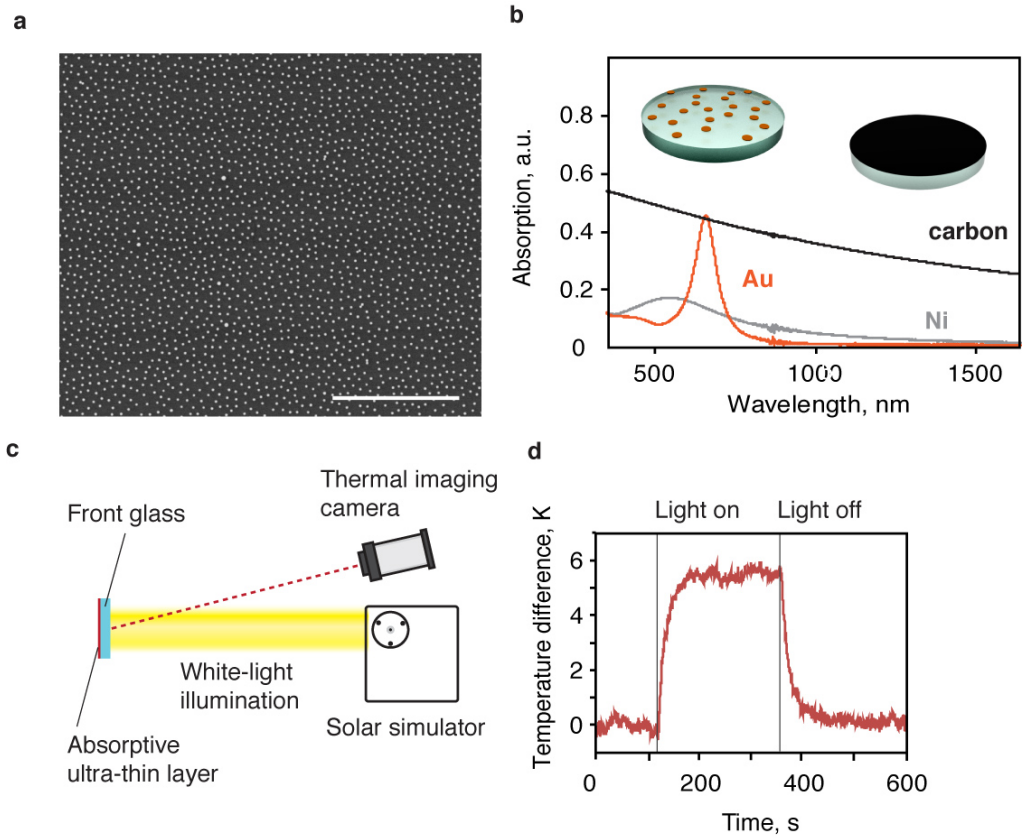

**Figure 1** (a) SEM micrograph of a plasmonic metasurface – amorphous arrangement of Ni nanodisks (scale bar – 5 µm); (b) Absorption spectra of Au and Ni nanodisks and of carbon thin film of the same thickness as the height of the plasmon particles. Insets schematically show both types of absorbing platforms; (c) Experimental setup for measuring temperature and (d) a typical temperature response.

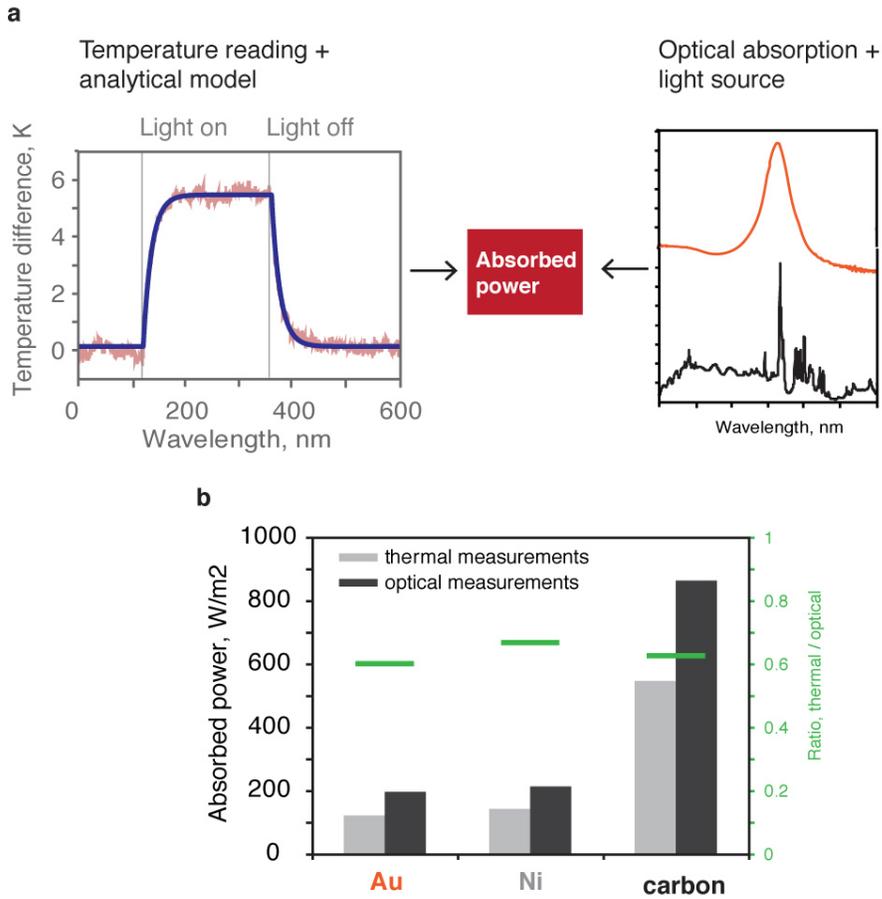

**Figure 2** (a) Schematic outline of the absorbed power measurement with two complimentary methods – temperature reading, fitted with analytical model (left) and optical absorption, combined with the emission spectrum of the light source (right). (b) Absorbed power from thermal (light grey) and optical (dark grey) measurements, green bars show the ratio between the two (right axis).

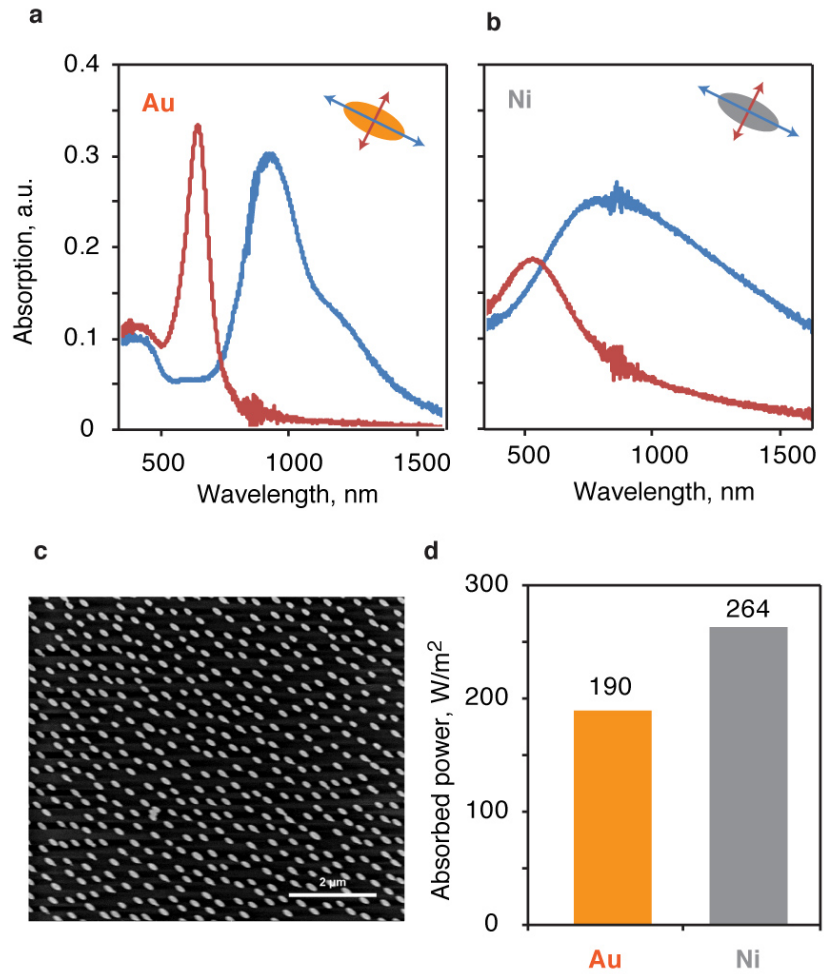

**Figure 3** Longitudinal (blue) and transversal (red) optical absorption in Au (a) and Ni (b) nanoellipses. Insets – schematics of the mutual orientation of nanoellipses axis and the polarization of the incoming light. (c) SEM of Ni amorphous metasurface array of nanoellipses. (d) Thermally measured absorbed power in Au (orange) and Ni (grey).

**Methods:**

*Sample preparation*

The plasmonic nanodisks and nanoellipses were prepared with the hole-mask colloidal lithography process as it is described elsewhere[9]. Polysterene particles with 110 nm diameter were used to fabricate the masks. The substrates were 0.21 mm thick circular borosilicate cover glasses with 25 mm diameter (D 263 M, Menzel-Gläser). The samples were entirely homogeneously covered by plasmonic nanoparticles.

The carbon thin films were evaporated on 0.5 mm thick fused silica substrates and annealed in Ar atmosphere at 800°C for 10 min.

*Optical absorption*

Optical absorption measurements were made using a Cary 5000 spectrophotometer [Varian] with an integrating sphere [DRA 2500, Varian] equipped with the small spot kit. The samples were positioned using the center mount sample holder which was rotated 9 degrees to include specular reflection. The baseline is recorded with an empty sample holder in place.

*ESEM*

SEM micrographs of the plasmonic nanostructures were obtained from the same samples as the rest of the measurements were made on using a Quanta 200 FEG ESEM, FEI. Particle sizes and number densities were determined from the micrographs using the software ImageJ 1.44o.

*Thermal absorption measurement*

The thermal measurement setup consists of a solar simulator [ss150, Sciencetech] as light source, a thermal camera [A20, FLIR] to monitor the sample temperature and a sample holder with the sample. The sample is positioned in the center of the solar simulator light beam at normal angle to ensure complete and homogeneous illumination. The intensity of the light is calibrated to 2 suns (2000 W/m$^2$) using a photodiode [Sciencetech]. The absorbing metamaterial side of the sample is facing away from the light source, i.e. is positioned at the backside of the sample. The thermal camera is directed towards the front side of the sample at an angle large enough for the IR-radiation, specularly reflected by the sample, to be that of the ambient environment and not from the camera itself or the solar simulator. The ambient environment has a known and constant emission as opposed to the thermal camera or the solar simulator, thus allowing reliable thermal measurements.

To reduce thermal contact, the sample holder balances the sample between two razorblade edges and a sharpened wire tip.

The emissivity of the glass surface was verified by heating the substrate above 50°C and measuring its temperature using a thin k-type thermocouple with its hot junction adhered to the substrate by Ag-glue and its cold junction kept at a controlled 0°C. At the same time the temperature of the substrate was measured with the thermal camera. As the temperature readings were equal, the correct emissivity can be recorded in the real experiments.

*Modeling*

The one-dimensional model is constructed using Comsol Multiphysics. It contains one domain, namely the substrate. This domain has the same length as the substrate thickness and the same material properties. Material propertied for the cover glasses are provided by the manufacturer. Fused silica properties were taken from the literature[23]. At both boundaries of the domain heat is transferred to the ambient using the software functions for thermal transport and thermal radiation. At one boundary a heat source is placed. The temperature is calculated at the opposite boundary. We fitted the model to our data by varying the power of the heat source. Our reading is the value of this parameter that gave the best fit to experimental data. Other ad hoc parameters in our model are ambient temperature and heat transfer coefficient. The ambient temperature is given by the equilibrium temperature of the sample in the dark for each measurement. The heat transfer coefficient is fixed for all samples with the same type of substrate.